\documentclass[preprint,12pt]{aastex}
\shortauthors{Gabel et al.}
\shorttitle{Variable UV Absorption in NGC 3783}

\begin{document}
\title{ The Ionized Gas and Nuclear Environment in NGC 3783
III. Detection of a Decreasing Radial Velocity in an Intrinsic UV Absorber\altaffilmark{1}}
\author{Jack R. Gabel\altaffilmark{2}, D. Michael Crenshaw\altaffilmark{3}, 
Steven B. Kraemer\altaffilmark{4}, W. N. Brandt\altaffilmark{5}, Ian M. George\altaffilmark{6,7},
Frederick W. Hamann\altaffilmark{8}, Mary Elizabeth Kaiser\altaffilmark{9}, Shai Kaspi\altaffilmark{5,10}, 
Gerard A. Kriss\altaffilmark{9,11}, Smita Mathur\altaffilmark{12},
Richard F. Mushotzky\altaffilmark{6}, Kirpal Nandra\altaffilmark{6,13}, Hagai Netzer\altaffilmark{10}, 
Bradley M. Peterson\altaffilmark{12}, Joseph C. Shields\altaffilmark{14},
T. J. Turner\altaffilmark{6,7}, \& Wei Zheng\altaffilmark{9}}
\altaffiltext{1}{Based on observations made with the NASA/ESA {\it Hubble 
Space Telescope}.  STScI is operated by the Association of Universities 
for Research in Astronomy, Inc. under NASA contract NAS~5-26555.}
\altaffiltext{2}{The Catholic University of America/IACS, NASA/Goddard Space Flight Center, Laboratory for Astronomy and Solar Physics, Code 681, Greenbelt, MD 20771.}
\altaffiltext{3}{Department of Physics and Astronomy, Georgia State University,
Atlanta, GA 30303.}
\altaffiltext{4}{The Catholic University of America, NASA/Goddard Space Flight Center, Laboratory for Astronomy and Solar Physics, Code 681, Greenbelt, MD 20771.}
\altaffiltext{5}{Department of Astronomy and Astrophysics, 525 Davey Laboratory, The 
Pennsylvania State University, University Park, PA 16802.}
\altaffiltext{6}{Laboratory for High Energy Astrophysics, NASA/Goddard Space Flight Center,
Code 662, Greenbelt, MD 20771.}
\altaffiltext{7}{Joint Center for Astrophysics, Physics Department, University of Maryland,
Baltimore County, 1000 Hilltop Circle, Baltimore, MD 21250.}
\altaffiltext{8}{Department of Astronomy, University of Florida, 211 Bryant Space Science Center,
Gainesville, FL, 32611-2055.}
\altaffiltext{9}{Center for Astrophysical Sciences, Department of Physics and Astronomy,
The Johns Hopkins University, Baltimore, MD 21218-2686.}
\altaffiltext{10}{School of Physics and Astronomy, Raymond and Beverly Sackler Faculty of
Exact Sciences, Tel-Aviv University, Tel-Aviv 69978, Israel.}
\altaffiltext{11}{Space Telescope Science Institute, 3700 San Martin Drive, Baltimore, MD 21218.}
\altaffiltext{12}{Department of Astronomy, Ohio State University, 140 West 18th Avenue,
Columbus, OH 43210-1173.}
\altaffiltext{13}{Universities Space Research Association, 7501 Forbes Boulevard, Suite 206, 
Seabrook, MD 20706-2253.}
\altaffiltext{14}{Department of Physics and Astronomy, Clippinger Research Labs 251B, Ohio
University, Athens,  OH 45701-2979.}

\begin{abstract}

     We report an intrinsic absorber with decreasing outflow velocity
in the Seyfert 1 galaxy NGC 3783.
This is the first detection of a change in radial velocity in an outflow
associated with a Seyfert galaxy.
These results are based on measurements from 18 observations 
with the Space Telescope Imaging Spectrograph aboard the 
{\it Hubble Space Telescope}, obtained between 2000 February and 2002 January.
In two intervals separated by $\sim$~13 and 9 months, 
the absorption lines in the kinematic component with highest outflow velocity exhibited
mean redward velocity shifts of $\sim$~35 and 55~km~s$^{-1}$, respectively.
The rate of velocity decrease was 2.2~$\pm$~0.6 times more rapid in the second interval.  
No variations in absorption velocities were detected in the other kinematic components.
We explore potential interpretations of the observed velocity shifts: radial deceleration 
of the UV absorber due to a change in either the speed or direction
of motion of the outflow, and the evolution of a continuous flow across our
line of sight to the emission source.
\end{abstract}


\section{Introduction}      

     Recent surveys with the {\it Hubble Space Telescope (HST)} 
and the {\it Far Ultraviolet Spectroscopic Explorer (FUSE)} have
shown intrinsic UV absorption is common in Seyfert 1 galaxies,
appearing in over half of the objects observed to date \citep{cren99,kris02}.  
In most cases, the absorption is blueshifted with respect to the 
systemic velocity of the host galaxy, 
indicating radial outflow from the active galactic nucleus (AGN).
There is a one-to-one correspondence between the detection of intrinsic
UV absorption and X-ray ``warm absorption" in Seyfert galaxies, indicating they 
are closely connected \citep{cren99}.
The mass outflow revealed by the blueshifted absorption may be a major
component of the energetics and circumnuclear environment in AGN 
\citep[e.g.,][]{math95,reyn97}.
Currently, the mechanism driving this mass outflow in AGN is not well understood.

    Variability of the intrinsic UV absorption appears to be common in Seyferts.
In the few objects that have multiple UV spectra at sufficiently high spectral resolution
for the accurate comparison of column densities, significant variations are detected
in all cases \citep{mara96,cren00a,krae02,cren03}.
In some extreme examples, entire kinematic components have appeared
and disappeared on roughly yearly timescales.
Strong evidence for column density variations ascribable both to ionization changes in response 
to the AGN continuum \citep{krae01a,krae02} and to changes in 
total absorption columns in our line of sight to the AGN, i.e., due to transverse 
motion \citep{krae01b,cren03}, have been found.

   One somewhat surprising finding has been the lack of variability in the
absorption kinematics. 
Although multi-epoch data are sparse, no detections of radial velocity variations
have been reported for any intrinsic absorbers in Seyfert galaxies, revealing stability 
on timescales of at least several years in some cases \citep[e.g.,][]{weym97,krae01b}.
Many of the extreme counterparts to 
Seyfert absorbers, the broad absorption line (BAL) systems seen in more
luminous AGN, have also shown this long-term stability \citep[e.g.,][]{barl93,weym02}.
Limits on radial acceleration in BALs have been used to assess models 
of the dynamical forces associated with their mass outflow \citep{folt87,weym97}.
Rare exceptions to this stability have been reported by 
\citet{hama97} and \citet{vilk01}, in which centroid radial velocity shifts were 
detected in relatively narrow absorption line systems in QSOs.
 
    The bright Seyfert 1 galaxy NGC 3783 has exhibited dramatic variability in 
its UV absorption spectrum.
In only four observations with the Goddard High Resolution Spectrograph (GHRS) and 
the Space Telescope Imaging Spectrograph (STIS) obtained approximately 1 -- 5 years
apart, new kinematic components appeared in three separate cases \citep{mara96,krae01b}.
No correlation was found between the column densities and continuum flux,
indicating these large-scale changes were primarily due to transverse
motion of the absorbers \citep{krae01b}.  In part because of this extreme variability, 
NGC 3783 was selected for an intensive multiwavelength monitoring campaign with 
{\it HST}, {\it FUSE}, and {\it Chandra}.
In this study, we concentrate on variability of the absorption kinematics based on 
analysis of the STIS spectra obtained during our monitoring campaign.  Analyses of 
the averaged X-ray and UV (STIS and {\it FUSE}) spectra are presented in 
\citet[][hereafter Paper I]{kasp02} and \citet[][hereafter Paper II]{gabe03}, respectively.  
Full treatment of the variability of the UV and X-ray absorption will be given in 
subsequent papers.

\section{Observations and Results}

     A total of 18 {\it HST}/STIS observations of the nucleus of NGC 3783
were obtained between 2000 February 27 and 2002 January 6 using the E140M 
medium resolution echelle grating.
This grating spans 1150~--~1730~ \AA\  with a spectral resolving power 
of $R \approx$~40,000, sampling the Ly$\alpha$, \ion{N}{5}~$\lambda\lambda$1238.8,1242.8,
\ion{C}{4}~$\lambda\lambda$1548.2,1550.8, and \ion{Si}{4}~$\lambda\lambda$1393.8,1402.8 resonance 
lines at a resolution of $\sim$8~km~s$^{-1}$.
Each visit consisted of two orbits and $\sim$4.9~ks of exposure time.
A log of the observations and full description of the data reduction are given in Paper~II.

   A detailed discussion of the UV absorption spectrum is presented in \citet{krae01b} 
and Paper~II.  
Here we review some features that are important for the subsequent analysis.
During our monitoring campaign, strong absorption was present in Ly$\alpha$, \ion{N}{5}, 
and \ion{C}{4} in three kinematic components, at centroid radial velocities of 
$v_r \approx -$1350, $-$725, and $-$550~km~s$^{-1}$ relative to the systemic
redshift of NGC 3783 \citep[$z =$~0.00976;][]{deva91}. 
For historical reasons, we refer to these as components 1, 3, and 2, respectively.
Weak absorption was tentatively detected in a fourth component at $-$1050~km~s$^{-1}$.
The \ion{Si}{4} doublet only appears in component 1.
Metastable \ion{C}{3} was detected in component~1 in the averaged spectrum, indicating
a high density in this absorber ($n_e \approx$ 10$^9$~cm$^{-3}$; Paper~II).
Absorption in these components first appeared in the following epochs:
component 2 in a 1994 January 16 GHRS spectrum of \ion{C}{4}, component 1 in a 1995 April 11 GHRS
spectrum of \ion{C}{4}, and components 3 and 4 in the first STIS spectrum (2000 February 27).

   We fit the intrinsic (i.e., unabsorbed) continuum plus emission-line fluxes in each STIS
spectrum following the procedure described in Paper~II.
Each spectrum was normalized by dividing by its fit. 
We compared the interstellar absorption lines in each spectrum to ensure that 
there were no wavelength offsets.
Specifically, we measured the centroid wavelengths of the Galactic \ion{Si}{4}~$\lambda$1393.8,
\ion{C}{4}~$\lambda$1548.2, and \ion{Fe}{2}~$\lambda$1608.5 lines in all spectra. 
For each interstellar line, the standard deviation of the centroids 
($\sim$~2 -- 4~km~s$^{-1}$) is less than a resolution element, indicating accurate
relative wavelength solutions for these observations.
Centroid radial velocities were then measured for each intrinsic absorption line.

    In Figure 1a, the \ion{N}{5}, \ion{C}{4}, and \ion{Si}{4} component 1 centroid velocities measured
in all STIS spectra are plotted, clearly showing the radial velocity has decreased during the 
monitoring campaign.
We identify three epochs exhibiting marked velocity variations:
the first observation obtained 2000 February 27, the intensive phase of the
campaign consisting of 13 observations obtained between 2001 February 25 -- April 23,
and the final observation obtained 2002 January 6.
Figure 1b show only the intensive monitoring observations (epoch 2).  Careful inspection
of Figure 1b reveals no correlations in the centroids measured for the three ions; the
fluctuations are likely due to the measurement uncertainties (see discussion below).

    Table~1 lists the component 1 velocity measurements for each of the three epochs
identified above.
Centroids and uncertainties quoted for epoch 2 are the means and standard deviations
of the 13 observations obtained during the intensive phase of the campaign.
These observations also represent a good statistical sample for determining the
random uncertainties associated with measurements in the individual spectra.
Thus, we adopted the standard deviations for the 1~$\sigma$ errors in epochs 1 and 3
as well, with the caveats that (1) if there are real variations in the centroid
velocity during the intensive phase of the campaign these are upper limits on 
the random uncertainties and (2) this does not account for any systematic errors that may
be associated with the measurements.
The radial decelerations listed in Table~1, $a_r$, assume a constant 
decrease in velocity between epochs. Uncertainties in these values were derived by
propagating the errors associated with the radial velocities. 
The average deceleration between epochs 2 and 3 was 2.2~$\pm$~0.6 times
greater than in the interval between epochs 1 and 2.
\ion{N}{5} shows some evidence for smaller velocity shifts than the other lines;
however, formally, the decelerations are equal within the 1~$\sigma$ uncertainties.
No radial velocity variations were detected in kinematic components 2 -- 4 within
the uncertainties in the measurements ($\Delta v_r \approx$~10--20~km~s$^{-1}$).
Interestingly, only the component with the highest outflow velocity is observed to be
decelerating.

     While a full treatment of the other absorption 
properties, i.e., absorption strengths, widths,
and covering factors, will be given in a later
study (Gabel et al., in preparation), we briefly
describe results here that have implications for
the interpretation of the radial velocity shift.
First, we note that although there were some 
variations in the absorption strengths and widths
during the monitoring campaign, there were no
large-scale changes in the three epochs that 
would indicate we are seeing different
components appearing and disappearing to
mimic a shift in velocity (although we may be seeing 
different regions of a continuous flow moving across
our line of sight, as discussed in \S 3.3).
This is evident from the overall similarity in the 
absorption spectra in Figure~2, which shows the profiles 
for each line in each of the three epochs.
There were moderate variations in the component~1 line widths during the STIS observations,
but these variations were complex; each line varied differently and
there were no apparent correlations with radial velocity ($FWHM$ are listed in Table~1).
For example, Figure~2 shows the mean \ion{C}{4} width in epoch~2
($FWHM =$150~km~s$^{-1}$) is narrower than
in epochs 1 and 3 ($FWHM =$203 and 194 $\pm$15 km~s$^{-1}$,
respectively), however the width varied
between 115 -- 190~km~s$^{-1}$ during the epoch 2 observations, on much shorter
timescales than the radial velocity shifts.
Furthermore, Table~1 shows the \ion{N}{5} widths were not correlated with
\ion{C}{4}.
On the other hand, the \ion{C}{4} profile in the 1995 GHRS spectrum is 
much broader ($FHWM =$500 $\pm$50 km~s$^{-1}$) than during
the STIS observations, as seen in Figure~1 in \citet{krae01b}.
Given the five year separation between the GHRS and STIS observations
and the dramatic difference in the profiles, we cannot rule out that
the absorption in the GHRS spectrum is unrelated to that observed with 
STIS; however, if it is, it indicates a significant evolution
of the absorber's structure over this interval.
There were no detectable changes in the line-of-sight covering factors
in the component~1 absorption lines.  This provides constraints on the evolution of the 
absorption/emission geometry between the three epochs.
Finally, we note there is evidence that the UV absorption in component~1 is
comprised of at least two subcomponents with different ionization structures.
This was initially suggested by \citet{krae01b} from photoionization 
modeling of the first STIS spectrum.  Also, the \ion{Si}{4} covering factor
in component~1 was found to be smaller than that of at least \ion{H}{1}
and \ion{N}{5} (Paper II).
Since all lines exhibit the velocity shift, this indicates both
subcomponents are decreasing in outflow velocity.

\section{Interpretation of Decreasing Radial Velocity}

\subsection{Deceleration of the Outflow?}

   One possible explanation for the observed decrease in radial velocity
is that the absorber is undergoing a bulk radial deceleration.
We first consider gravity as the source of the required inward radial force.
The simplest case, gravitational deceleration of a ballistically ejected
absorber, is excluded by the {\it increased} deceleration
detected in the second interval (see Table~1), since the radial distance between the absorber and
central mass has increased between epochs (by $\sim$3$\times$10$^{15}$~cm).
This requires the presence of an outward radial force that has decreased
relative to gravity between the two intervals.
Since the other kinematic components in NGC 3783 are not decelerating, this
adds the further requirement that the nature of the radial forces for these
components differs from those of component~1, e.g., they may have reached
sufficient distances from the central mass that they are near terminal velocity.
Using the average deceleration over the second interval listed in Table~1 and 
the black hole mass for NGC~3783  \citep[$M_{BH} =$9$\times$10$^{6}$~M$_{\sun}$;][]{onke02}, 
the absorber would be $\leq$~9$\times$10$^{16}$~cm
from the central mass if gravity is the decelerating force (the upper limit
is for a negligible outward force during this interval).

    An alternate decelerating mechanism could be an interaction of the 
absorber with other gas in the vicinity of the AGN. 
This has been invoked to explain observations of deceleration 
in the narrow line region (NLR) gas in several Seyfert galaxies
because the onset of deceleration in these objects is at distances well 
beyond the gravitational influence of the central mass \citep{cren00c,cren00a,ruiz01}.
In the context of intrinsic absorbers, this could be the signature of an
interaction between different phases in the outflow, as predicted
in many dynamical models \citep[e.g.,][]{bege91,krol01,ever03}.
The hydrodynamic drag force is proportional to $\Delta v^2 n_i / N_a$,
where $\Delta v$ is the
difference in velocity between the absorber and the ambient medium,
$n_i$ is the density of the ambient medium, and $N_a$ is the column density
of the absorber \citep{bege91}.
In this framework, a twofold increase in $n_i / N_a$ is required in the second interval
to account for the observed increased deceleration.
However, a potential problem for this model is the absence of any signatures
of the interaction in the absorption spectrum.
The deceleration of the NGC 3783 absorber is much greater than that observed
in the NLR gas, requiring a drag force (per unit mass) that is four orders of
magnitude greater ($F_d / m \approx$ 0.05 cm~s$^{-2}$).
Changes in the ionization structure appear in the decelerating
region of the NLR gas, presumably due to shocks \citep{krae00,ceci02},
whereas no such changes are detected in the absorption spectrum in NGC 3783, as
discussed above.

   A third possible mechanism is deceleration of the outflow due to mass loading.
In this scenario, additional mass is swept up by a momentum conserving
flow, resulting in a decreased outflow velocity.
This is believed to be the dominant effect in the deceleration of supernova
remnants \citep[e.g.,][and references therein]{dyso02}.  
As an example of how an outflow's mass could be increased, we note
the dynamical models of \citet{ever03} predict material from the accretion disk
is continually lifted by a disk-driven wind and added to the flow.  
Since the deceleration will depend on the rate of increase in the flow's mass, 
the increased deceleration observed in the second interval requires that mass was
accumulated more rapidly than during the first interval.

\subsection{Directional Shift in the Outflow?}

     Another interpretation of the observed decrease in blueshifted
radial velocity is that, rather than a bulk radial deceleration as
described above, we are seeing a change in the radial component of the
velocity vector with respect to our line of sight to the AGN.
Figure~3 illustrates this schematically, depicting the absorber as 
a cloud moving along a curved path across our sight line to the 
background emission source.  
In this simple model, the velocity of the cloud, represented
with thick filled arrows in Figure~3, has a constant amplitude
along the trajectory, but the component that is directed radially in our line of 
sight (open arrows in the $x_r$ direction) decreases due to the changing direction
of the motion.
The absorber occults $\sim$30\% of the BLR (Paper~II), which is demarcated with 
dashed lines in Figure~3.

    The transverse component of motion of the absorber
(i.e., along $x_T$ in Figure~3) is an integral part of this model, leading
to the motion of the absorber into and eventually out of our view.
A rough lower limit on the transverse velocity $v_T$ can be derived
using the transverse size of the absorber (1$\times$10$^{16}$~cm; Paper~II)
and an upper limit on the time required for the absorber to move into our
line of sight.
As a conservative estimate of the latter, we use the interval between the 
1994 January GHRS observation and the first STIS observation, since it is
uncertain whether the absorption in the 1995 April GHRS spectrum is from the
same absorber as that in the STIS data (see discussion in \S 2).
The result is $v_T \geq$~542~km~s$^{-1}$, which, in combination with the
transverse size of the BLR \citep[1.9$\times$10$^{16}$~cm;][]{onke02}, 
gives a rough upper limit on the total time that some portion of the absorber 
will occult the BLR, $t \leq$~17~yrs. 
This is strictly an upper limit since (a) we have only an upper limit on the 
interval over which the absorption appeared and (b) $v_T$ is increasing, 
as illustrated in Figure~3.
Since the continuum source is certainly much smaller
than the BLR, at some point in the absorber's path it will 
no longer occult the continuum source but will still occult the same fraction
of the BLR (see Paper~II for a full discussion of the covering factors).
Thus, this model predicts the covering factors, and therefore absorption depths,
could change substantially at a later epoch, depending on the specific
geometry of the BLR and continuum sources.
Since the covering factor was not observed to change between the three epochs
in our STIS campaign as described in \S 2, this implies the absorber has not yet
moved out of our line of sight to the continuum source.
We note that the rate of change in radial velocity will be determined by
the curvature of the absorber's trajectory and will not necessarily be
constant. This may provide a natural explanation for the more rapid decrease
in radial velocity observed in the second interval of the observations.

     A number of mechanisms could produce the curved motion depicted in Figure~3.
For example, models positing the absorber is driven off the accretion
disk by radiation pressure from the central source \citep[e.g.,][]{murr95,prog00}
or by centrifugal forces along magnetic field lines in the disk 
\citep[e.g.,][]{blan82,bott00,ever03} all predict the outflow will follow a curved 
trajectory above the disk, exhibiting a changing angle $\theta$ with respect to the 
perpendicular to the disk.
Additionally, these models predict the outflow's path will be helical due to 
rotation about the central mass, and thus its motion will be curved in the plane
parallel to the disk.

\subsection{Evolution of a Continuous Flow?}

     If the component~1 absorber is a continuous outflow, the centroid
velocity could vary due to the motion of different regions of the 
flow across our line of sight to the emission source.
In this scenario, the absorber could be either a continuous stream of gas or a continuous
``spray" of small clouds.
An example is shown in Figure~4, in which a collimated flow, dileneated with thick solid lines, 
intersects our line of sight to the emission 
source obliquely.  The mean angle of intersection in the flow, $\theta$, determines 
the centroid radial velocity.
Due to its finite opening angle, $\Delta \theta$, the flow spans a range of projected radial 
velocities, giving the observed width of the absorption feature, $v \times (cos(\theta
 + \Delta \theta/2) - cos(\theta -  \Delta \theta/2))$.
Two epochs are depicted in Figure~4, showing how changes in the flow geometry lead to 
variations in the absorption kinematics.
Centroid velocity variations could be produced by either an overall change in the flow's direction
or inhomogeneities in the flow within a given opening angle that result in a change in $\theta$.
In constrast to the models discussed in the previous sections, the observed velocity shift is 
due to the motion of different gas elements, with different 
projected radial velocities, into our line of sight
rather than the change in speed or direction of motion of any
given gas element.

\section{Conclusions}

   STIS spectra obtained as part of our monitoring campaign of NGC 3783
reveal a decreasing blueshifted radial velocity in one of its
intrinsic absorption components.
This is the first detection of a variable radial velocity in an outflow associated
with a Seyfert galaxy.
In two intervals separated by $\sim$~13 and 9 months, 
the absorption lines in the kinematic component with highest outflow velocity exhibited
mean redward velocity shifts of $\sim$~35 and 55~km~s$^{-1}$, respectively.
The rate of decrease in the second interval was twice as large as in the first.
The absorption widths in this component varied moderately during these observations.
These variations were complex, fluctuating on short timescales compared to the velocity shifts
and showing no apparent correlation with the variations in radial velocity.
No variations in absorption velocities were detected in the other kinematic components.

     We have explored various interpretations of the observed velocity shift.
The similarity in spectra at all epochs indicates it is not due to changes in 
the relative strengths of multiple subcomponents, and it thus represents a {\it bona
fide} decrease in radial velocity.  
If it represents bulk deceleration due to a decrease in radial speed, 
we consider gravity, interaction with an ambient medium, and mass loading as 
possible deceleration mechanisms and derive special requirements on the
nature of the radial forces.
Alternatively, it could be due to a directional shift in the motion of the absorber
with respect to our line of sight to the background emission sources, or
a result of different regions in a continuous flow crossing our sightline.

     J. R. G., D. M. C., and S. B. K. acknowledge support from NASA grant 
HST-GO-08606.13-A.  W.N.B acknowledges CXC grant GO1-2103 and NASA LTSA grant 
NAG5-13035.  F.H. acknowledges NSF grant AST 99-84040. We thank the anonymous
referee for very helpful comments.

\clearpage

\clearpage

\begin{deluxetable}{lllllllllll}
\tabletypesize{\scriptsize}
\tablewidth{0pt}
\tablecaption{Variable Kinematics in Component 1 \label{tbl-1}}
\tablehead{
\colhead{Epoch} & \colhead{Julian Date} & \multicolumn{3}{c}{N V} &
\multicolumn{3}{c}{C IV} & \multicolumn{3}{c}{Si IV\tablenotemark{a}}\\
\colhead{} & \colhead{} & \colhead{$v_r$\tablenotemark{b}} & \colhead{$a_r$\tablenotemark{c}} & \colhead{$FWHM$}  & 
\colhead{$v_r$\tablenotemark{b}} & \colhead{$a_r$\tablenotemark{c}} & \colhead{$FWHM$} & \colhead{$v_r$\tablenotemark{b}} & \colhead{$a_r$\tablenotemark{c}} & \colhead{$FWHM$}\\
\colhead{} & \colhead{(2,450,000 +)} & \colhead{(km s$^{-1}$)} & \colhead{(10$^{-6}$ km s$^{-2}$)} & 
\colhead{(km s$^{-1}$)} & \colhead{(km s$^{-1}$)} &
\colhead{(10$^{-6}$ km s$^{-2}$)} & \colhead{(km s$^{-1}$)} & \colhead{(km s$^{-1}$)} & 
\colhead{(10$^{-6}$ km s$^{-2}$)} & \colhead{(km s$^{-1}$)}}
\startdata
1 & 1602 & $-$1334$\pm$4 & \nodata & 212$\pm$15 & $-$1352$\pm$8 & \nodata & 203$\pm$15 & $-$1374$\pm$13 & \nodata & 175$\pm$25  \\
2\tablenotemark{d} & 1966--2023 & $-$1307$\pm$4 & $-$0.8$\pm$0.2 & 140--212 & $-$1318$\pm$8 & $-$1.0$\pm$0.3 & 115--190 & $-$1333$\pm$13 & $-$1.2$\pm$0.5 & 100--145  \\
3 & 2281 & $-$1261$\pm$4 & $-$1.8$\pm$0.3 & 150$\pm$15 & $-$1256$\pm$8 & $-$2.5$\pm$0.5 & 194$\pm$15  & $-$1280$\pm$13 & $-$2.2$\pm$0.8 & 140$\pm$25 \\
\tablenotetext{a}{Average values for the two lines of the Si IV doublet.}
\tablenotetext{b}{Centroid velocity of component 1 relative to the systemic velocity.}
\tablenotetext{c}{Average radial deceleration since previous epoch.}
\tablenotetext{d}{$v_r$ and $a_r$ are mean values and $FWHM$ are range of values measured
during intensive phase of monitoring.}
\enddata
\end{deluxetable}

\clearpage

\begin{figure}
\epsscale{0.7}
\plotone{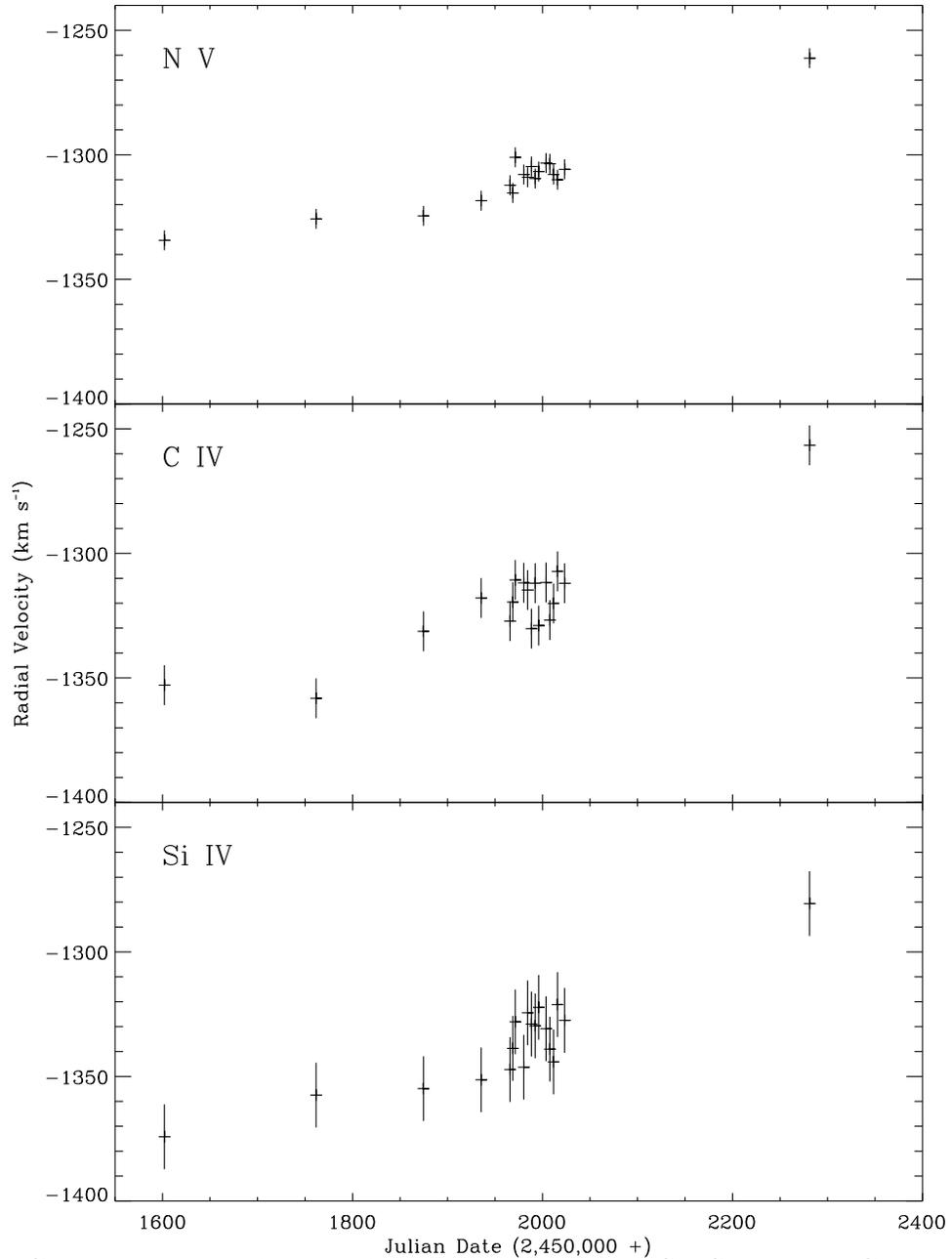}
\caption{Component~1 radial velocities measured in the STIS spectra.
Centroid velocities and error bars are plotted for \ion{N}{5}~$\lambda$1238.8 (top), 
\ion{C}{4}~$\lambda$1548.2 (middle) and the average of the two \ion{Si}{4} doublet
lines (bottom) as a function of Julian date. 
Measurements from all observations are shown in (1a); only observations obtained 
during the intensive monitoring are shown in (1b).
\label{fig1a}}
\end{figure}

\addtocounter{figure}{-1}

\begin{figure}
\epsscale{0.7}
\plotone{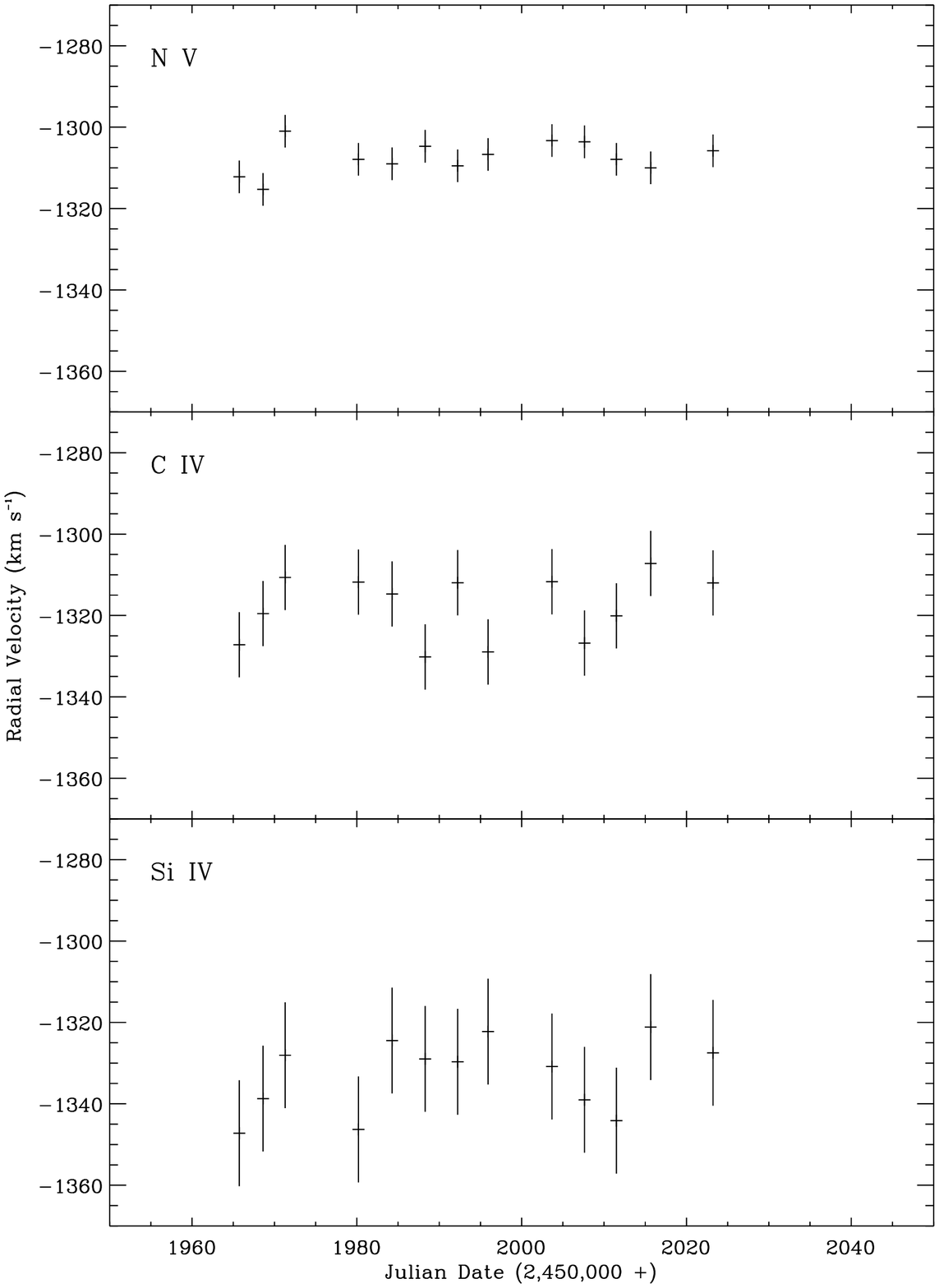}
\caption{\label{fig1b}}
\end{figure}

\clearpage

\begin{figure}
\epsscale{0.7}
\plotone{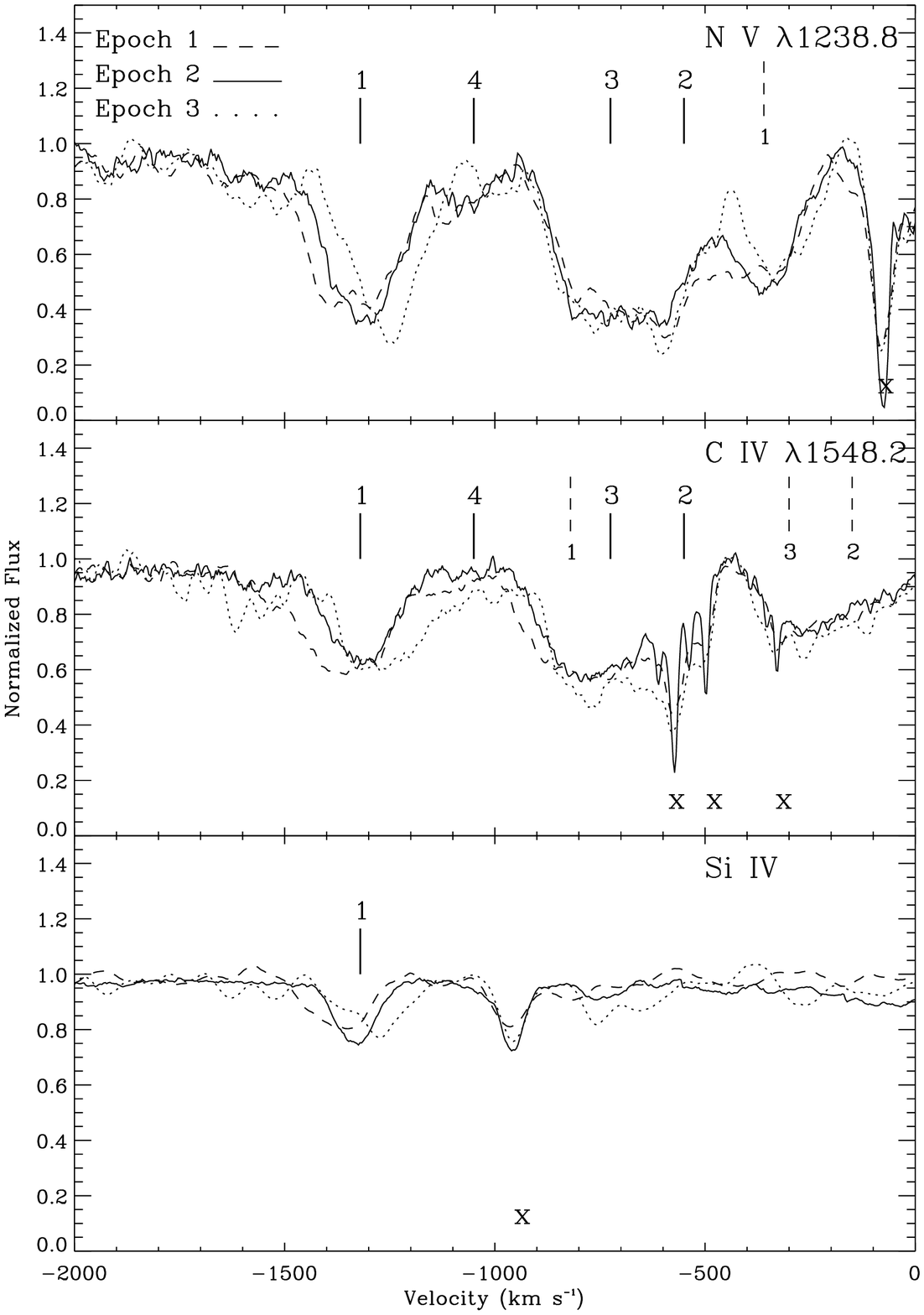}
\caption{Normalized \ion{N}{5}, \ion{C}{4}, and \ion{Si}{4} absorption profiles for three epochs
showing the decrease in the component 1 radial velocity.
Profiles are plotted in velocity space with respect to the short wavelength lines of
the doublets in the rest frame of NGC 3783.  
Epoch 1 is the first observation (dashed line), epoch 2 is the average of 13 observations 
obtained in the intensive phase of the campaign (solid line), and epoch 3 is the final 
observation (dotted line).  
All intrinsic absorption features are marked with tick marks above the spectra and numbered
following the conventions described in the text: lower, thick
marks identify the short wavelength doublet members; upper, thin marks identify the
long wavelength members.
The \ion{Si}{4} profile is the mean of the two doublet lines.
Interstellar absorption features are marked with an ``x" below each spectrum. 
Epochs 1 and 3 have been smoothed for clarity, leading to shallower interstellar lines compared
to the composite spectrum of epoch 2. \label{fig2}}
\end{figure}

\begin{figure}
\epsscale{0.7}
\plotone{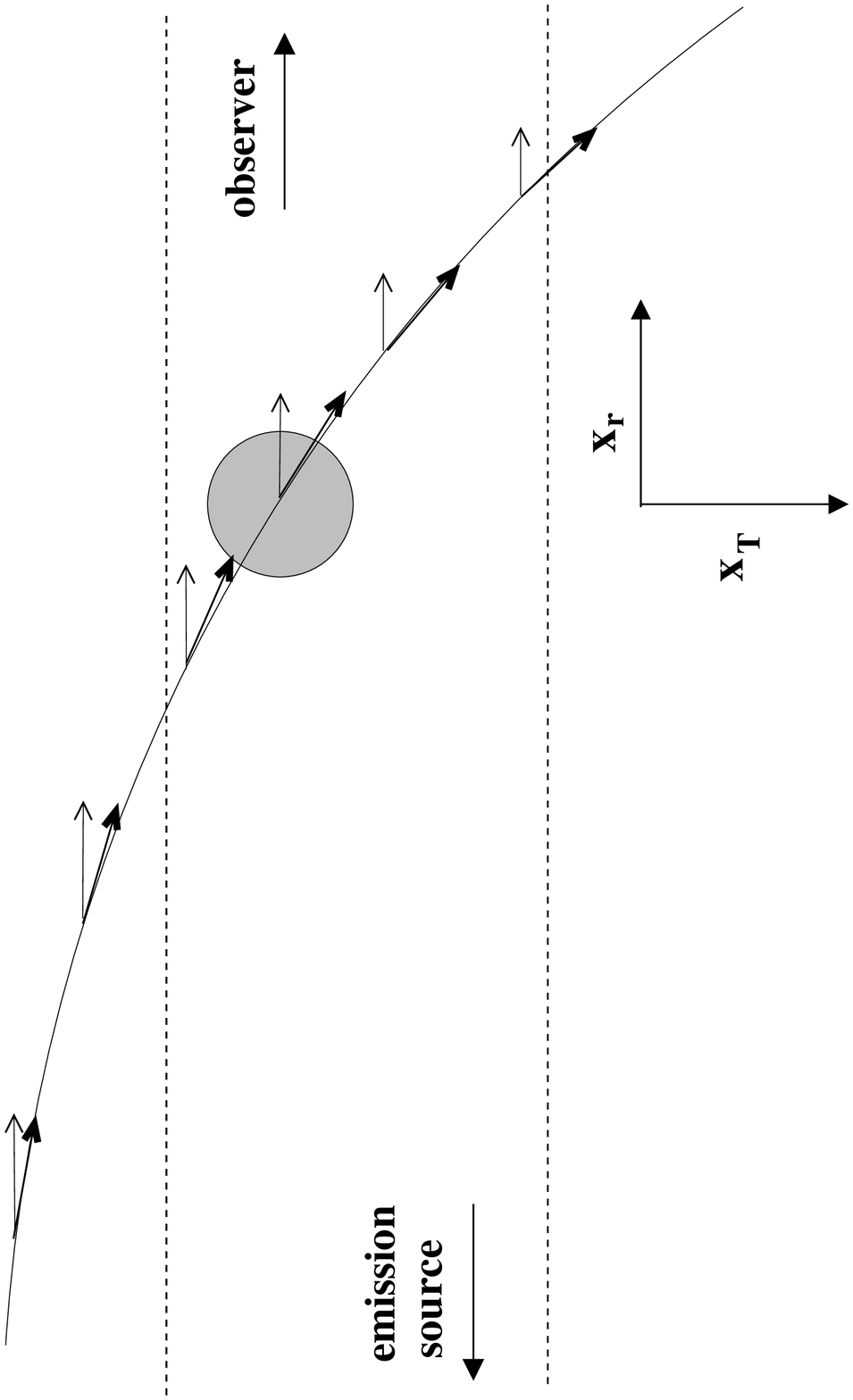}
\caption{Schematic representation of a simple model in which the curved path of an absorption
cloud leads to the observed decrease in radial velocity.
The dashed lines outline the transverse size of the BLR projected on the sky.
The coordinate system, relative to the observer, is shown at the top right.
The absorber is depicted as a shaded circle that occults $\sim$30\% of the BLR and
moves along a curved path at constant velocity (thick, filled arrows). 
The radial component of the velocity (thin, open arrows) is seen to decrease at
different points on the clouds trajectory. \label{fig3}}
\end{figure}

\begin{figure}
\epsscale{0.8}
\plotone{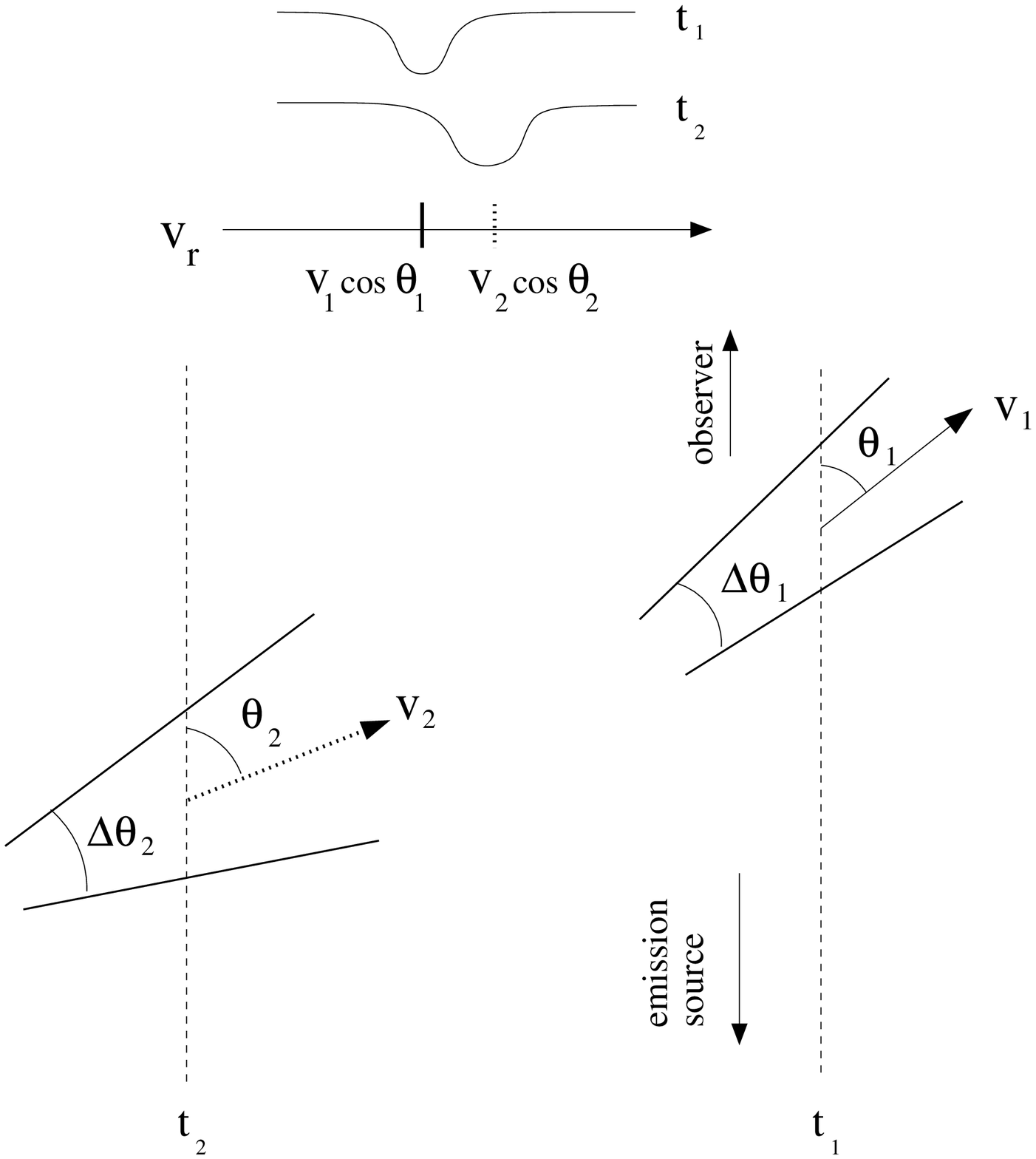}
\caption{Schematic demonstrating how kinematic variability could arise in a continuous flow.
Two epochs are shown, t$_1$ and t$_2$.  In each epoch, a collimated outflow, 
delineated with thick solid lines, intersects
the observers line of sight (dashed lines) at an oblique angle. 
Due to the opening angle of the flow, a range of angles, $\Delta \theta$,
is spanned in the line of sight giving the absorption line width. 
The mean intersection angle, $\theta$, determines the centroid.
The resulting variability in absorption profiles due to the geometrical differences
in different regions of the flow is shown at the top. \label{fig4}}
\end{figure}


\clearpage

\begin{thebibliography}{}
\bibitem[Barlow (1993)]{barl93}Barlow, T. A. 1993, Ph.D. thesis Univ. of California at San Diego
\bibitem[Begelman et al.(1991)]{bege91}Begelman, M. C. de Kool, M., \& Sikora, M. 1991, \apj, 382, 416
\bibitem[Blandford \& Payne (1982)]{blan82}Blandford, R. D., \& Payne, D. G. 1982, \mnras, 199, 883
\bibitem[Bottorff et al.(2000)]{bott00}Bottorff, M. C., Korista, K. T., \& Shlosman, I. 2000, \apj, 537, 134
\bibitem[Castor et al.(1976)]{cast76}Castor, J. I., Abbott, D. C., \& Klein, R. I. 1976, \apj, 195, 157
\bibitem[Cecil et al.(2002)]{ceci02}Cecil, G., Doptia, M. A., Groves, B., Wilson, A. S., Ferruit, P., Pecontal, E., \& Binette, L. 
2002, \apj, 568, 627
\bibitem[Crenshaw et al.(1999)]{cren99}Crenshaw, D. M., Kraemer, S. B., Boggess, A., Maran, S. P., Mushotzky, R. F., 
\& Wu, C.-C. 1999, \apj, 516, 750
\bibitem[Crenshaw \& Kraemer (2000)]{cren00c}Crenshaw, D. M., \& Kraemer, S. B. 2000, \apj, 532, L101
\bibitem[Crenshaw et al.(2000a)]{cren00a}Crenshaw, D. M., et al. 2000a, \apj, 545, L27
\bibitem[Crenshaw et al.(2000b)]{cren00b}Crenshaw, D. M., et al. 2000b, \aj, 120, 1731
\bibitem[Crenshaw et al.(2003)]{cren03}Crenshaw, D. M., et al. 2003, \apj, in press
\bibitem[de Vaucouleurs et al.(1991)]{deva91}de Vaucouleurs, G., de Vaucouleurs, A., Corwin, H. G., Buta, R. J., Paturel, 
G., \& Fouque, P.  1991 Third Reference Catalogue of Bright Galaxies (Springer-Verlag: New York)
\bibitem[Dyson et al.(2002)]{dyso02}Dyson, J. E., Arthur, S. J., \& Hartquist, T. W. 2002, A\&A, 390, 1063
\bibitem[Everett (2003)]{ever03}Everett, J. E. 2003, \apj, in press, astro-ph/0212421
\bibitem[Foltz et al.(1987)]{folt87}Foltz, C. B., Weymann, R. J., Morris, S. L., \& Turnshek, D. A. 1987, \apj, 317, 450
\bibitem[Gabel et al.(2003)]{gabe03}Gabel, J. R., et al. 2003, \apj, 583, 178 (Paper II)
\bibitem[Hamann et al.(1997)]{hama97}Hamann, F., Barlow, T. A., Cohen, R. D., Junkkarinen, V., \& 
Burbidge, E. M. 1997, in ASP Conf. Ser. 128, Mass Ejection from Active Galactic Nuclei, ed. N. Arav,
I. Shlosman, \& R. J. Weymann (San Francisco: ASP), 19
\bibitem[Kaspi et al.(2002)]{kasp02}Kaspi, S., et al.\ 2002, \apj, 574, 643 (Paper I)
\bibitem[Kraemer \& Crenshaw (2000)]{krae00}Kraemer, S. B. \& Crenshaw, D. M. 2000, \apj, 544, 763
\bibitem[Kraemer et al.(2001a)]{krae01a}Kraemer, S. B., et al. 2001a, \apj, 551, 671
\bibitem[Kraemer et al.(2001b)]{krae01b}Kraemer, S. B., Crenshaw, D. M., Gabel J. R. 2001b, \apj, 557, 30
\bibitem[Kraemer et al.(2002)]{krae02}Kraemer, S. B., et al. 2002, \apj, 577, 98
\bibitem[Kriss (2002)]{kris02}Kriss, G. A.  2002, in ASP Conf. Ser. 255, Mass Outflow in Active Galactic Nuclei: 
New Perspectives, ed. D. M. Crenshaw, S. B. Kraemer, \& I. M. George (San Francisco: ASP), 69
\bibitem[Krolik \& Kriss (2001)]{krol01}Krolik, J. H. \& Kriss, G. A. 2001, \apj, 561, 684
\bibitem[Maran et al.(1996)]{mara96}Maran, S. P., Crenshaw, D. M., Mushotzky, R. F., Reichert, G. A., Carpenter, K. G.,
Smith, A. M., Hutchings, J. B., \& Weymann, R. J. 1996, \apj, 465, 733
\bibitem[Mathur et al.(1995)]{math95}Mathur, S., Elvis, M. \& Wilkes, B. 1995, \apj, 452, 230
\bibitem[Murray et al.(1995)]{murr95}Murray, N., Chiang, J., Grossman, S. A., \& Voit, G. M. 1995, \apj, 451, 498
\bibitem[Onken \& Peterson (2002)]{onke02}Onken, C. A., \& Peterson, B. M.  2002, \apj, 572, 746
\bibitem[Proga et al.(2000)]{prog00}Proga, D. Stone, J. M., \& Kallman, T. R. 2000, \apj, 543, 686
\bibitem[Reynolds (1997)]{reyn97}Reynolds, C. S. 1997, \mnras, 286, 513
\bibitem[Ruiz et al.(2001)]{ruiz01}Ruiz, J. R., et al. 2001, \aj, 122, 2961
\bibitem[Vilkoviskij \& Irwin (2001)]{vilk01}Vilkoviskij, E. Y., \& Irwin, M. J. 2001, \mnras,
321, 4
\bibitem[Weymann et al.(1997)]{weym97}Weymann, R. J., Morris, S. L., Gray, M. E., \& Hutchings, J. B. 1997, \apj, 483, 717
\bibitem[Weymann (2002)]{weym02}Weymann, R. J. 2002, in ASP Conf. Ser. 255, Mass Outflow in Active Galactic Nuclei: 
New Perspectives, ed. D. M. Crenshaw, S. B. Kraemer, \& I. M. George (San Francisco: ASP), 329
\end{thebibliography}
\end{document}